\documentclass[aps,prd,reprint,preprintnumbers,showpacs,nofootinbib,superscriptaddress,longbibliography]{revtex4-2}
\usepackage{natbib}

\usepackage[utf8]{inputenc}
\usepackage[T1]{fontenc}
\usepackage[english]{babel}

\usepackage{amsmath,amsfonts,amssymb}
\usepackage{mathtools}
\usepackage{mathrsfs}
\usepackage{tensor}
\usepackage{accents}
\allowdisplaybreaks

\usepackage{graphicx}
\usepackage[dvipsnames]{xcolor}

\usepackage{etoolbox}
\usepackage{xspace}

\setcitestyle{numbers,comma,square}

\usepackage{hyperref}
\hypersetup{colorlinks=true, linkcolor=Blue, citecolor=Blue, filecolor=Blue, urlcolor=Blue}
\usepackage[capitalize,nameinlink]{cleveref}
\crefname{equation}{Eq.}{Eqs.}
\Crefname{equation}{Eq.}{Eqs.}
\crefname{figure}{Fig.}{Figs.}
\Crefname{figure}{Fig.}{Figs.}
\crefname{section}{Sec.}{Secs.}
\Crefname{section}{Sec.}{Secs.}
\creflabelformat{equation}{#2(#1)#3}
\AddToHook{cmd/appendix/before}{\crefalias{section}{appendix}}

\makeatletter
\renewcommand{\paragraph}{%
\@startsection{paragraph}{4}%
{\z@}{1.21ex \@plus 1ex \@minus .2ex}{0.9em}%
{\normalfont\normalsize\bfseries}%
}
\usepackage{titlesec}
\newrobustcmd{\pea}[1]{\emph{#1}\textbf{.\ \ \ ---}}
\titleformat{\paragraph}[runin]{\normalfont\normalsize\bfseries}{\emph\theparagraph}{1em}{\pea}
\titleformat{\section}[block]{\normalfont\bfseries\centering}{\MakeUppercase\thesection}{1em}{\MakeUppercase}
\makeatother

\makeatletter \def\switch@array{}\makeatother

\newcommand{\MyEquiv}{\coloneqq}
\newcommand{\smallbinom}[2]{{\textstyle\binom{#1}{#2}}}

\def\a{\alpha}\def\b{\beta}\def\g{\gamma}\def\d{\delta}\def\D{\Delta}
\def\e{\epsilon}\def\vf{\varphi}
\let\polishl\l\DeclareUnicodeCharacter{0142}{\polishl}\def\l{\lambda}
\let\PolishL\L\DeclareUnicodeCharacter{0141}{\PolishL}\def\L{\Lambda}
\def\m{\mu}\def\n{\nu}\def\s{\sigma}
\def\h{\eta}

\def\pr{\partial}\def\prd{\partial \cdot}\def\pe{\prime}

\newcommand{\be}{\begin{equation}}
\newcommand{\ee}{\end{equation}}
\newcommand{\bea}{\begin{equation}\begin{aligned}}
\newcommand{\eea}{\end{aligned}\end{equation}}

\begin{document}

\title{\emph{Higher spin Lagrangians from higher derivative diffeomorphisms}}

\author{Will Barker}
\email{barker@fzu.cz}
\affiliation{Central European Institute for Cosmology and Fundamental Physics, Institute of Physics of the Czech Academy of Sciences, Na Slovance 1999/2, 182 00 Prague 8, Czechia}

\author{Dario Francia}
\email{dario.francia@uniroma3.it, dario.francia@roma3.infn.it}
\affiliation{Roma Tre University and INFN Sezione di Roma Tre, via della Vasca Navale, 84 I-00146 Roma, Italy}

\author{Carlo Marzo}
\email{carlo.marzo@kbfi.ee}
\affiliation{Laboratory for High Energy and Computational Physics, NICPB, R\"{a}vala 10, Tallinn 10143, Estonia}

\author{Alessandro Santoni}
\email{asantoni@uc.cl}
\affiliation{Institut f\"ur Theoretische Physik, Technische Universit\"at Wien, Wiedner Hauptstrasse 8--10, A-1040 Vienna, Austria}
 \affiliation{Facultad de F\'isica, Pontificia Universidad Cat\'olica de Chile, Vicu\~{n}a Mackenna 4860, Santiago, Chile}

\begin{abstract}
The covariant description of massless particles of arbitrary spin typically employs symmetric tensors of rank~$s$ and rests on a local symmetry carried by symmetric tensor parameters of rank~$s-1$, suitably generalizing the~$U(1)$ transformations of Maxwell's theory and the diffeomorphisms underlying general relativity. Here we show that Fronsdal's action is actually uniquely identified by the weaker requirement of invariance under higher-derivative gauge transformations driven by a vector parameter. This observation may hint to a tighter, if unexpected, connection between higher-spin symmetry and diffeomorphisms.
\end{abstract}

\maketitle

\paragraph*{Introduction} \label{intro}

In this letter we aim to identify the minimal amount of gauge symmetry required to uniquely select an appropriate kinetic tensor for massless fields of a given spin. By \emph{appropriate} we mean that the equations defined by setting to zero the kinetic tensor propagate the degrees of freedom pertaining to the corresponding massless particle. 

Following~\cite{Fierz:1939zz,Weinberg:1995mt,Fronsdal:1978rb, Curtright:1979uz}, the natural covariant encoding of the degrees of freedom of massless particles of spin~$s$ is given by rank-$s$ symmetric Lorentz tensors~$\vf_{\m_1 \ldots \m_s}$ to be interpreted as gauge potentials subject to the transformation
\begin{equation} \label{gauge}
\vf_{\m_1 \ldots \m_s} \rightarrow \vf_{\m_1 \ldots \m_s} + \pr_{\m_1} \, \e_{\m_2 \ldots \m_s + \, \ldots} \, ,
\end{equation} %
involving a rank-$(s-1)$ symmetric parameter. In particular in~\cite{Fronsdal:1978rb}, Fronsdal found the first general Lagrangian suitable for the description of the massless, spin-$s$ degrees of freedom encoded in~\cref{gauge}, building on the zero-mass limit of the corresponding massive Lagrangians proposed by Singh and Hagen in~\cite{Singh:1974qz}. Fronsdal's Lagrangian is indeed invariant under~\cref{gauge}, although with a parameter constrained to be traceless. In a local Lagrangian theory without additional fields, this is the largest amount of gauge symmetry allowed for the description of massless irreducible spin-$s$ degrees of freedom. 

 Gauge symmetry typically plays two related, yet distinct, roles. On the one hand, it is a group-theoretical feature of the covariant description of $iso(D-2)$ massless representations with finite spin \cite{Weinberg:1995mt}. On the other hand, it serves as a guideline to build off-shell kinetic tensors and Lagrangians for the corresponding degrees of freedom, via the Noether procedure (see e.g. \cite{Berends:1984rq, Barnich:1993vg, Bekaert:2002uh, Manvelyan:2010jr, Joung:2012fv, Francia:2016weg, Ponomarev:2022vjb}). The standard assumption is that the gauge parameters entering the two settings are tensors of the same rank.

In this letter we question this identification. In particular, we show that it is possible to still uniquely identify the Fronsdal kinetic tensor under the request of invariance under a {\it reduced} gauge symmetry, where the parameter in~\cref{gauge} is itself a multiple gradient of a tensor of lower rank. Our main result is to show that it suffices indeed to ask for invariance under higher-derivative linearised diffeomorphisms, i.e., gauge transformations like~\cref{gauge} but with parameters of the form
\begin{equation} \label{vector_param}
\e_{\m_1 \ldots \m_{s-1}} = \pr_{\m_1} \, \ldots \pr_{\m_{s-2}} \e_{\m_{s-1}}\, + \, \ldots \, .
\end{equation}
The simplification is sizeable: it means that, in order to fix the form of the Fronsdal kinetic tensor for any $s$, one only needs to ask for invariance under $D$ independent directions, instead of ${\cal{O}} (s^{D-2})$ (related to the number of components of a traceless parameter of rank $s-1$) as usually assumed. In addition, it suggests a somewhat unexpected role of diffeomorphisms in ruling higher-spin dynamics.
The possibility that this might be the case was put forward in~\cite{Barker:2025xzd} by means of an explicit example for spin three.

We also prove that under the strongest possible reduction, employing the~$s-$th gradient of a scalar parameter, the condition of gauge invariance does \emph{not} suffice to uniquely select the Fronsdal form. This shows in particular that no reduction is possible for the spin-two case, and that in this sense our result characterizes a phenomenon peculiar to higher spins. 

We start by reviewing how to construct free higher-spin Lagrangians from the bottom-up, enforcing (as far as possible) invariance under~\cref{gauge}. We then proceed to implement the same ideas for the case of a reduced gauge symmetry. Details about the notation and technical material concerning the analysis of the reduction to a scalar parameter are collected in~\cref{notation,Failure}, and in the supplement~\cite{Supplement}.

%%%
\paragraph*{Higher-spin Lagrangians} \label{hsplagr}
%%%
Given a gauge potential~$\vf \MyEquiv \vf_{\m_1 \ldots \m_s}$ subject to~\cref{gauge},  a systematic way to derive free Lagrangians for massless fields is to first build a gauge invariant kinetic tensor~$E$, defining the equations of motion in the form~$E=0$, and then construct a Lagrangian whose equations imply the vanishing of~$E$, possibly via some intermediate steps. The equations~$E=0$ should imply the propagation of only  physical degrees of freedom at zero mass. This requires in particular that they must imply the conditions\footnote{We follow the notation and the computational rules outlined in~\cref{notation}.}~$\Box \vf = 0$ and~$\prd \vf  = 0$, together with analogous conditions on the gauge parameter, thus ensuring the propagation of massless degrees of freedom and no longitudinal modes. Additional conditions on the traces of~$\vf$ and~$\e$ determine the spectrum of the resulting theory. We look for kinetic tensors of order two in derivatives.  Assuming locality and the absence of auxiliary fields,  this program admits two solutions (up to field redefinitions).

The kinetic tensor~$E$ has to take the form
\begin{equation} \label{E}
E = \Box \vf + \D (\vf) ,
\end{equation}
where~$\D (\vf)$ is some local tensor whose gauge variation under~\cref{gauge} has to compensate~$\d \Box \vf = \Box \pr \e$. It is easy to realize that~$\D (\vf)$ has to be of the form 
\begin{equation}
\D (\vf) = - \pr \prd \vf + \ldots .
\end{equation}
In this sense, off the mass-shell (i.e. for~$p^2 \neq 0$) any covariant theory has to include the Maxwell-like block
\begin{equation} \label{ML}
M \MyEquiv \Box \vf - \pr \prd \vf ,
\end{equation}
whose overall variation is~$ \d M = - 2 \pr^2 \prd \e .~$ One first possibility is to allow in~\cref{gauge} only parameters satisfying the transversality condition~$ \prd \e = 0$ and to identify~\cref{ML} as the kinetic tensor of the theory~\cite{CampoleoniFrancia2013}. Alternatively, one can further compensate the variation of~$M$. To this end, the only possible local term available employs the trace of~$\vf$ and builds the Fronsdal tensor~\cite{Fronsdal:1978rb}
\begin{equation} \label{F}
F \MyEquiv \Box \vf - \pr \prd \vf + \pr^2 \vf^{\pe} ,
\end{equation}
embodying a higher-spin generalisation of the linearised Ricci tensor on flat backgrounds, whose gauge variation involves the trace of the gauge parameter~$\d F = 3 \pr^3 \e^{\pe}$. The latter, however, cannot be compensated in any local way,\footnote{In particular, higher traces of~$\vf$ can enter only in combination with the metric tensor~$\h_{\m\n}$ and for this reason they cannot contribute to compensating~$\d F$. Various options to remove the trace conditions, either employing additional fields or foregoing locality, were discussed e.g. in~\cite{Francia:2002aa, Bekaert:2003az, Bekaert:2003uc, Francia:2005bu, Francia:2007qt, Buchbinder:2007ak, Bittencourt:2025roa}.} thus providing a rationale to Fronsdal's trace constraint~$\e^{\pe} = 0$.  Self-adjointness of the operator building~$M$ allows one a straightforward construction of the corresponding Lagrangian 
\begin{equation} \label{Mlagr}
{\cal{L}} = \frac{1}{2} \vf \cdot M \, ,
\end{equation}
while in the Fronsdal case it is necessary to combine~$F$ with its trace so as to get the gauge invariant Lagrangian
\begin{subequations}
\begin{align}
{\cal{L}} &= \frac{1}{2} \vf \cdot \Big(F - \frac{1}{2} \h F^{\pe}\Big)\, , \label{Flagr} \\
F^{\, \pe} &\MyEquiv 2 \Box \vf^{\, \pe} + \pr \, \prd \vf^{\, \pe} -2 \prd \prd \vf\, , \label{Ftrace}
\end{align}
\end{subequations}
whose equations of motion eventually imply~$F = 0$, under the further condition that the gauge-invariant double-trace of~$\vf$ be set to zero from the outset:\footnote{Under this condition~\cref{Ftrace} identifies the trace of~$F$.}~$\vf^{ \pe \pe} = 0$.  Fronsdal's equations describe the free propagation of a single massless particle of spin~$s$, and in particular imply that, on shell,~$\vf^{\, \pe} = 0$. By contrast, the particle spectrum associated with the Maxwell-like Lagrangian in~\cref{Mlagr} depends on the possible trace constraints that one can assume on fields and parameters, on top of the transversality condition~$\prd \e = 0$. In particular, by taking both~$\vf$ and~$\e$ to be traceless from the outset, one recovers an off-shell, gauge-fixed version of the Fronsdal theory, with the same spectrum~\cite{Skvortsov:2007kz}. Weaker trace conditions, on the other hand, allow for the propagation of additional massless particles of lower spin. For fields and parameters subject to the condition of~$k$th tracelessness,~$\vf^{\, [k]} = 0 = \e^{\, [k]}$, the equations of motion derived from~\cref{Mlagr} describe a unitary spectrum of particles of spin~$s$,~$s-2$,~$\ldots$,~$s- 2(k-1)$~\cite{Francia:2016weg}. Assuming no trace conditions at all one obtains a fully reducible theory involving a set of massless particles of spin~$s$,~$s-2$,~$s-4$,~$\ldots$ down to~$s=1$ or~$s=0$~\cite{CampoleoniFrancia2013}. This is the same spectrum described by the triplets emerging from tensionless strings~\cite{Bengtsson:1986ys, Henneaux:1987cp, Francia:2002pt, Sagnotti:2003qa, Fotopoulos:2008ka}, to which~\cref{Mlagr} is indeed strictly related~\cite{Francia:2011qa, Francia2010}. 

These considerations show that the form of the free Lagrangian for arbitrary spin is identified by the requirement of invariance under~\cref{gauge}, whilst the dichotomy between the two solutions in~\cref{Mlagr} and~\cref{Flagr} depend on the type of constraints assumed on the gauge parameters. 

In this letter, we focus on the single-particle description of massless higher spins, and in this respect we make reference to Fronsdal's solution, which displays a larger gauge symmetry than its irreducible Maxwell-like counterpart (the latter implementing a transverse and traceless parameter while the former only assuming the parameter to be traceless). Our goal is to show that by requiring a much \emph{weaker} form of gauge invariance, only involving a multiple gradient of a \emph{vector} parameter, it is still possible to identify the Fronsdal Lagrangian as the only possible gauge invariant solution.

%%%
\paragraph*{Reduced symmetry: general setup} \label{hspfromdiff} 
%%%
In this section we reconsider the construction of a gauge invariant Lagrangian for a rank-$s$ tensor, assuming a restricted form of gauge variation where the parameter is itself a multiple gradient of a lower-rank tensor. More precisely, we consider the following setup
\begin{equation} \label{epsilon_k}
\begin{gathered}
\d \vf = \pr \hat{\e}_{s-1} \, , \quad \hat{\e}_{s-1} = \sum_{k=0}^{[\frac{s-1}{2}]} \a_k \h^k \e_{s-1}^{[k]} \, , \\
\e_{s-1} = \pr^k \e_{s-1-k} \, ,
\end{gathered}
\end{equation}
where~$\hat{\e}_{s-1}$ is a traceless parameter of rank~$s-1$,~$\e_{s-1}$ is an unconstrained rank-$(s-1)$ tensor, while the coefficients
\begin{equation} \label{alphas}
\a_k \MyEquiv \frac{(-1)^k}{\prod_{l=1}^k [D + 2(s-l-2)]} \, , \quad \a_0 \MyEquiv 1 \, ,
\end{equation}
implement in \eqref{epsilon_k} its traceless projection, and~$\e_{s-1-k}$ is an unconstrained parameter of rank~$s-1-k$. We assume the parameter~$\hat{\e}_{s-1}$ to be traceless, as we want the Fronsdal solution to be included. For the ensuing manipulations, it is useful to compute the variations of  divergence and trace of~$\vf$ under~\cref{epsilon_k}:
\begin{subequations}
\begin{align}
\d \, \prd \vf &= \sum_n \a_n \Big[\h^n \big[\Box (\pr^k \e_{s-1-k})^{[n]} + \pr \prd (\pr^k \e_{s-1-k})^{[n]} \big] \nonumber \\
& + \h^{n-1} 2 \pr^2 (\pr^k \e_{s-1-k})^{[n]} \Big] , \label{deltadiv_k} \\
\d \, \vf^{\, \pe} &= \sum_n \a_n \Big[\h^n \big[2 \prd (\pr^k \e_{s-1-k})^{[n]} \nonumber \\
& + \pr (\pr^k \e_{s-1-k})^{[n+1]} \big] \nonumber \\
& + \h^{n-1} [D+2(s-n-1)] \pr (\pr^k \e_{s-1-k})^{[n]} \Big] . \label{deltatrace_k}
\end{align}
\end{subequations}

For~$k > 0$ quantities of the form~$\e_{s-1} = \pr^k \e_{s-1-k}~$ identify a strict subset of the Fronsdal parameters, while the sum of all possible quadratic monomials involving up to two derivatives of~$\vf$ appearing in a candidate Lagrangian involves a number of arbitrary coefficients. Thus, a priori, one might expect that requiring the Lagrangian to be gauge invariant, up to total derivatives, admits~$k$-dependent solutions. 

In the spirit of the discussion of the previous section, we approach the problem from the simpler perspective of constructing kinetic tensors~$E_k$ of the form in~\cref{E}, providing gauge invariant completions of~$\Box \vf$ under~\cref{epsilon_k}. Assuming locality and the absence of additional fields, one finds that the general form of~$E_k$ involves five possible tensor structures besides~$\Box \vf$:
\begin{equation}
 \begin{split} \label{E_k}
 E_k \, (\l_1, \ldots, \l_5) & = \Box \vf + \l_1 \pr \prd \vf + \l_2 \pr^{\, 2} \vf^{ \pe} \\
 & \hspace{-20pt} + \h \big[ \l_3 \Box \vf^{\, \pe} + \l_4 \pr \prd \vf^{ \pe} + \l_5 \prd \prd \vf\big].
 \end{split}
\end{equation}
We do not include higher traces of~$\vf$ that, being identically gauge invariant for traceless parameters, cannot contribute to the compensation of~$\d \Box \vf$. In particular, we are not setting the double trace of $\vf$ to zero. Our task is to find the possible values of~$\l_1$,~$\ldots$,~$\l_5$ such as ensure that~$\d E_k \, (\l_1, \ldots, \l_5) = 0$ under~\cref{epsilon_k}. In the following we focus on the two extremal cases of reductions to vector and scalar parameters, respectively, since these constitute the `strongest' possible reductions in~\cref{epsilon_k}. All the intermediate cases,~$0< k \leq s -3$, can be dealt with in a similar fashion.

%%%
 \paragraph*{Fronsdal from vector reduction} \label{vector_par}
%%%

In the third of~\cref{epsilon_k} let us first choose~$\e_{s-1 - k}$ to be a vector, i.e. take~$k = s-2$ so that
\begin{equation} \label{epsilon_k_vector}
\e_{s-1} = \pr^{s-2} \, \e_1\, ,
\end{equation}
or, upon displaying indices for better clarity,
\begin{equation} \label{epsilon_indices}
\e_{\m_1 \ldots \m_{s-1}} = \pr_{(\m_1} \, \ldots \pr_{\m_{s-2}} \e_{\m_{s-1})} \, ,
\end{equation}
so that the gauge variation of a symmetric tensor is of the form
\begin{equation} \label{delta_phi_indices}
\d \vf_{\m_1 \ldots \m_s} = s \, \pr_{(\m_1} \, \pr_{\, \m_2} \dots \pr_{\m_{s-1}} \e_{\m_s)} \, + \ldots,
\end{equation}
where in~\cref{epsilon_indices,delta_phi_indices} the parentheses are meant to enforce symmetrization involving the minimal number of terms, without normalization factors, while the dots refer to the terms enforcing the traceless projection displayed in the second relation of \eqref{epsilon_k}.

By computing the corresponding variations of the first three terms in~\cref{E_k} one obtains
\begin{subequations}
\begin{align}
\d \, \Box \vf &= \sum_n \a_n \h^n \Big[\g_{-1} \Box^{n+1} \pr^{\, \g_{-1}} \e_1 \nonumber \\
& \quad \hspace{20pt} + 2n \g_0 \Box^n \pr^{\, \g_0} \prd \e_1 \Big] , \label{deltabox_k} \\
\d \, \pr \, \prd \vf &= \sum_n \a_n \h^{n-1} \Big[\g_{-1} \g_0 \g_1\Box^n \pr^{\, \g_1} \e_1 \nonumber \\
& \quad \hspace{20pt} + 2n \g_0 \g_1 \g_2 \Box^{n-1} \pr^{\, \g_2} \prd \e_1\Big] \nonumber \\
& \quad + \sum_n \a_n \h^n \Big[\g_{-1}^2 \Box^{n+1} \pr^{\, \g_{-1}} \e_1 \nonumber\\
& \quad \hspace{20pt} + \g_0\big(\g_{-1} + 2n\g_0\big) \Box^n \pr^{\, \g_0} \prd \e_1\Big] , \label{deltadediv_k} \\
\d \, \pr^{\, 2} \vf^{\, \pe} &= \sum_n \a_n \h^{n-1}\big[D + 2\big(\g_{-1}+n\big)\big] \nonumber \\
& \quad \times \Big[\g_{-1} \smallbinom{\g_1}{2}\Box^n \pr^{\, \g_1} \e_1 \nonumber \\
& \quad \hspace{20pt} + 2n \g_0 \smallbinom{\g_2}{2}\Box^{n-1} \pr^{\, \g_2} \prd \e_1 \Big] \nonumber \\
& \quad + \sum_n \a_n \h^n \Big[\g_{-1} \smallbinom{\g_{-1}}{2} \Box^{n+1} \pr^{\, \g_{-1}} \e_1 \nonumber \\
& \quad \hspace{20pt} + 2 \big(\g_{-1} + n\g_0\big) \smallbinom{\g_0}{2}\Box^n \pr^{\, \g_0} \prd \e_1 \Big] , \label{deltadesquaretrace_k}
\end{align}
\end{subequations}
where the coefficients~$\alpha_n$ were defined in~\cref{alphas}, whilst~$N \choose l$ denotes the binomial coefficient and we further define~$\g_l \MyEquiv s -2n + l$. Altogether, for the variation of~\cref{E_k} we get
\begin{align}
\d E_{s-2} &= \Box \pr^{s-1} \e_1 \Bigg[s-1 + \l_1 \bigg[(s-1)^2 - \frac{6 \smallbinom{s-1}{3}}{D+2s-6}\bigg] \nonumber \\
& \hspace{10pt} + \l_2 \bigg[(s-1)\smallbinom{s-1}{2} - \frac{3 (D + 2s -4)\smallbinom{s-1}{3}}{D+2s-6} \bigg]\Bigg] \nonumber \\
& + \pr^s \prd \e_1 \Bigg[\l_1\bigg[s (s-1) - \frac{12 \smallbinom{s}{3}}{D+2s-6}\bigg] \nonumber \\
& \hspace{10pt} + \l_2 \bigg[2 (s-1) \smallbinom{s}{2} - \frac{6(D + 2s -4)\smallbinom{s}{3}}{D+2s-6} \bigg]\Bigg] \nonumber \\
& + {\cal O} (\h) , \label{delta_Ek_vector}
\end{align}
where~${\cal O} (\h)$ denotes terms proportional to the spacetime metric and its higher powers. Such terms must cancel out independently, and in general they depend on all five coefficients~$\l_1$,~$\ldots$,~$\l_5$.

In~\cref{delta_Ek_vector} one can appreciate that, taking the contributions from the terms~${\cal O} (\h)$ into account, the number of independent structures that have to mutually compensate threatens to overdetermine the system of coefficients for the~$\l_1$,~$\ldots$,~$\l_5$. We know, however, that the system is compatible, since the Fronsdal solution exists: it corresponds to
\begin{equation} \label{fronsdal_lambdas}
\l_1 = -1, \quad \l_2 = 1,
\end{equation}
with the additional terms building a tensor proportional to 
$F^{\, \pe}$ as defined in~\cref{Ftrace}. Our goal is to investigate whether one can find kinetic tensors other than~\cref{F} that are still gauge invariant  under~\cref{delta_phi_indices} but that won't be invariant under the full higher-spin symmetry of~\cref{gauge}.

The relevant observation at this stage is that requiring~$\l_1$ and~$\l_2$ to cancel the terms in~\cref{delta_Ek_vector} not involving~$\h$ factors, already provides a system of two independent equations, whose {\it unique} solution is~\cref{fronsdal_lambdas}, for any $s$ and any $D$, thus building the Fronsdal kinetic tensor in~\cref{F}. In other words what we actually find is that, even after reducing the independent gauge transformations in \eqref{gauge} as drastically as to being higher-derivatives of a vector parameter, still the Fronsdal tensor gets uniquely identified as the basic gauge invariant combination of two-derivative local tensors in $\vf$. This is our main observation. In agreement with the general program outlined in the previous section, the  gauge-invariant equation $F=0$ describes the local degrees of freedom of a massless particle of spin $s$, and in particular implies  the double-trace condition $\vf^{\, \pe \pe} = 0$ \cite{Francia:2011qa}.  

 As for the remaining terms, all contributions in the~${\cal O} (\h)$ part involving the coefficients~$\l_1$ and~$\l_2$ also cancel out, and one is left with the condition that the remaining contributions of~${\cal O} (\h)$, depending on the coefficients~$\l_3$,~$\l_4$ and~$\l_5$, should sum up to zero autonomously. This is tantamount to finding coefficients such that 
\begin{equation}
\d \Big[\l_3 \Box \vf^{\, \pe} + \l_4 \pr \prd \vf^{\, \pe} + \l_5 \prd\prd\vf\Big] = 0 \, .
\end{equation}
Barring an overall rescaling, it is not hard to recognize that one builds in this way the counterpart of the system in~\cref{delta_Ek_vector}, up to terms~${\cal O} (\h^2)$, whose unique solution yields a tensor  proportional to~\cref{Ftrace}
\begin{equation} \label{lambda345_solution}
\l_3 = c, \quad \l_4 = \frac{c}{2}, \quad \l_5 = -c \, ,
\end{equation}
for some constant~$c$, rather then a new independent gauge-invariant quantity. 

The double trace condition $\vf^{\pe \pe} = 0$, as well as the value of~$c$, get fixed by the requirement that the equation $F=0$  be a consequence of the equations obtained from the Lagrangian ${\cal{L}} = \frac{1}{2} \vf \cdot (F - \frac{1}{2} \h F^{ \pe})$, whose gauge invariance relies on the identity
\begin{equation}
\prd F - \frac{1}{2} \pr F^{\, \pe} \equiv 0\, ,
\end{equation}
holding indeed for doubly-traceless fields \cite{Francia:2002aa}.
This completes our proof that~\cref{Flagr} is the unique solution for a Lagrangian invariant under~\cref{epsilon_k} in the particular case of reduction to a vector gauge parameter.

\paragraph*{Scalar reduction} \label{scalar_par}

\begin{figure*}[t]
 \centering
 \includegraphics[width=\textwidth]{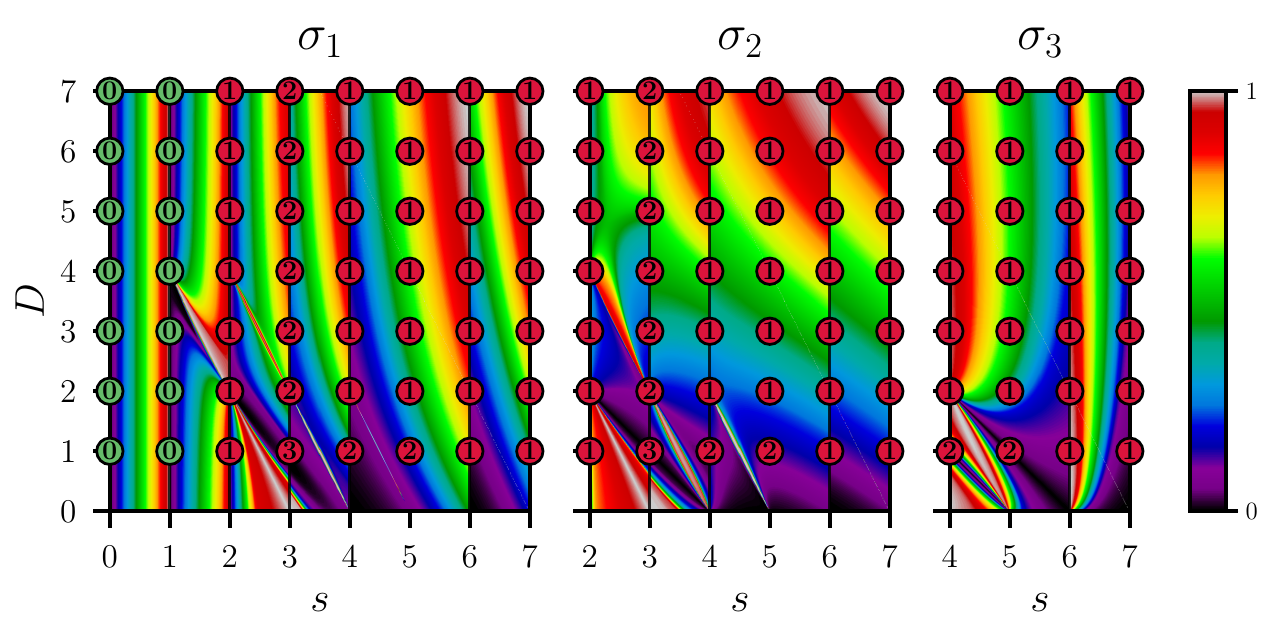}
 \caption{For spin~$s$ and dimension~$D$, the circled numbers indicate the degree to which the scalar reduction \emph{fails} to motivate the unique Fronsdal solution, expressed as the number~$N(s,D)$ of remaining free parameters beyond Fronsdal. This number is determined in~\cref{NCalc} by the rank of the linear system that encodes scalar-reduced gauge invariance in~\cref{ScalarFunction}. The rank at any point corresponds to the number of non-zero singular values~$\s_1$,~$\s_2$ and~$\s_3$, represented here using a heatmap in arbitrary units, where~$\s_2$ emerges only for~$s\geq2$, and~$\s_3$ emerges only for~$s\geq 4$. The analytic continuation in~$s$ and~$D$ reveals how the plane is punctuated by singular value roots passing through~$D=1$ at~$s=3$,~$s=4$ and~$s=5$. These anomalous points have even larger~$N(s,D)$, and so fail more spectacularly. No noteworthy structure exists for~$s>7$ or~$D>7$.}
 \label{fig:singular_values_francia}
\end{figure*}

The reduction to a scalar parameter~$\e$, such that 
\begin{equation}
\e_{\m_1 \ldots \m_{s-1}} = \pr_{(\m_1} \, \ldots \pr_{\m_{s-1})} \e \, ,
\end{equation}
can be discussed in a similar fashion but delivers a different outcome: in this case the Fronsdal solution {\it is not} uniquely selected. Let us consider the variation in~\cref{epsilon_k} with~$k=s-1$ and look for coefficients~$\l_1$,~$\ldots$,~$\l_5$ such as to render the kinetic tensor in~\cref{E_k} gauge invariant. The main difference with respect to the previous case is that imposing gauge invariance in the sector~${\cal O} (\h^0)$ does not suffice to fix~$\l_1$ and~$\l_2$ to their Fronsdal values, and one has to solve for all coefficients in the generic sectors~${\cal O} (\h^n)$. Indeed, at~${\cal O} (\h^n)$ the variations of the tensors entering~\cref{E_k} are
\begin{subequations}
\begin{align}
\d \, \Box \vf^{ (n)} &= \a_n (s-2n) , \label{deltabox_n} \\
\d \, \pr \, \prd \vf^{(n)} &= \a_{n+1} (s-2n-2) (s-2n-1) \nonumber \\
& \hspace{30pt} \times (s-2n) + \a_n (s-2n)^2, \label{deltadediv_n} \\
\d \, \pr^{\, 2} \vf^{\, \pe \, (n) } &= \Big[\a_{n+1} (s-2n-2) \nonumber \\
& \hspace{30pt} \times [D+2 (s-n-2)] \nonumber \\
& \hspace{20pt} + \a_n (s-2n)\Big] \smallbinom{s-2n}{2}, \label{deltadesquaretrace_n} \\
\d \, \h \, \Box \, \vf^{\, \pe \, (n) } &= \a_n\, n \, (s-2n) [D+2 (s-n-1)] \nonumber\\
& \hspace{20pt} + \a_{n-1} (s-2n+2), \label{deltaetaboxtrace_n} \\
\d \, \h \, \pr \prd \vf^{\, \pe \, (n) } &= n \, (s-2n) \Big[\a_{n-1} (s-2n+2) \nonumber \\
& \hspace{20pt} + \a_n (s-2n) [D+3s -4n -3] \nonumber \\
& \hspace{20pt} + \a_{n+1} (s-2n-2)(s-2n-1) \nonumber \\
& \hspace{30pt} \times [D+2 (s-n-2)] \Big], \label{deltadeddottracephi_n} \\
\d \, \h \, \prd \prd \vf^{\, (n) } &= n \Big[\a_{n-1} (s-2n+2) \nonumber \\
& \hspace{20pt} + \a_n (s-2n) (2s - 4n +1) \nonumber \\
& \hspace{20pt} + \a_{n+1} (s-2n -2) (s-2n -1) \nonumber \\
& \hspace{30pt} \times (s-2n)\Big] , \label{deltaddotddotphi_n}
\end{align}
\end{subequations}
from which one can appreciate in particular that gauge invariance at~${\cal O} (\h^0)$ only implies one condition for the two coefficients~$\l_1$ and~$\l_2$:
\begin{align}
\d E_{s-1} & = \frac{s \Box \pr^s \e}{D + 2(s-3)} \Big[[D + 2(s-3)] \big(1 + s\l_1 + \smallbinom{s}{2} \l_2\big) \nonumber \\
& \hspace{10pt} - \smallbinom{s-1}{2} \big(2\l_1 + [D + 2(s-2)] \l_2\big)\Big] + {\cal O (\h)} \, .\label{deltaEkorder0}
\end{align}
A few comments are in order:
\begin{itemize}
\item \Cref{deltaEkorder0} marks a difference with the vector reduction, where at the same order one can already uniquely identify the structure of the Fronsdal tensor~$F$.
 \item In particular,~\cref{deltaEkorder0} shows that for any spin $s$ the reduction  to a scalar parameter does not identify uniquely the kinetic tensor. In this sense, the reduction as a general mechanism is a phenomenon peculiar to higher spins, since it is only for $s\geq3$ that one can reduce (up to vector parameters) the original symmetry and still uniquely identify the correct kinetic tensor. Indeed for~$s=2$,  the second term vanishes, so that one can always find kinetic tensors other than the linearised Ricci tensor that are gauge invariant under~$\d h_{\m\n} = \pr{_\m} \pr_{\n} \e$. Given a generic sample from this set of solutions, the corresponding equations of motion would not describe a single massless particle of spin two and would generically propagate ghosts.
 \item Let us notice, however, that for arbitrary spin~$s>2$, upon further asking that the coefficients~$\l_1$ and~$\l_2$ do not depend on~$D$ one recovers the Fronsdal solution uniquely (while also getting for free spin-independence of the coefficients).  This additional requirement would not change our conclusions for the spin-two case, for which  the dependence on~$D$ in~\cref{deltaEkorder0} gets factorised, so that all the spurious gauge invariant solutions are still generated by~$D$-independent coefficients.
 \end{itemize}
As we show in~\cref{Failure,fig:singular_values_francia}, even beyond the terms~${\cal O} (\h^0)$ our analysis shows that one can always find invariant tensors other than Fronsdal, with coefficients that in general are dependent on the spacetime dimension.

\paragraph*{Concluding remarks} \label{comments}

In this work we proved that the structure of the Fronsdal kinetic tensor is rigidly determined by asking invariance under a much smaller number of local symmetries than usually considered, a number that only depends on the spacetime dimension and not on the spin. Put differently, we showed that the theory of a rank-$s$ potential $\vf$ asked to be invariant under a local symmetry as small as in ~\cref{delta_phi_indices} is anyway due to propagate a massless spin$-s$ particle and nothing else, although in principle the vast majority of the longitudinal components in $\vf$ are not demanded to be pure gauge from the beginning. The full higher-spin symmetry of the Fronsdal theory, in this perspective, might be interpreted as emerging, or accidental. 

In our construction it is important to impose that the reduced parameter is traceless. This allows the recovery of the Fronsdal solution, for which the `enhanced' symmetry under a full rank-$(s-1)$ traceless parameter grants the elimination of all unphysical polarizations. Removing the traceless projection from~\cref{epsilon_k} would produce solutions devoid of this crucial property. For instance, the vector reduction for a spin-four gauge field performed on an  {\it unconstrained} parameter of spin three would produce the unique gauge invariant kinetic tensor
\be\label{E4}
 E_4 = \Box \vf - \frac{1}{2} \pr \prd \vf + \frac{1}{6} \pr^2 \vf^{\, \pe}.
\ee
For the kinetic tensor in~\cref{E4} there is no hidden enhancement of the symmetry: thus, the expectation is that the corresponding equation~$E_4=0$ would not propagate the correct degrees of freedom and the spectrum would contain longitudinal modes. 

We focused on the metric-like formulation of bosonic higher-spin theories in flat space. One could explore the counterpart of the mechanism that ensures the sufficiency of the reduced symmetry on (A)dS backgrounds, for fermions, and in the frame-like approach. (For the latter see \cite{Ponomarev:2022vjb} for a recent review.)

For the Maxwell-like theory of~\cite{Skvortsov:2007kz} and ~\cite{CampoleoniFrancia2013} one cannot try a reduction of the type in~\cref{epsilon_k}, given the transversality of the gauge parameter assumed in that context. Thus, in a sense, the construction that we explored in this letter, based on longitudinal parameters, provides some sort of complementary route to the Maxwell-like case that exploits purely transverse parameters.
 
In both cases, the indication that emerges is that the full gauge symmetry encoded in~\cref{gauge}, even when restricted to traceless parameters, is redundant to the goal of deriving a consistent Lagrangian for massless fields with spin $s>2$.\footnote{In a similar spirit, an alternative formulation for single-spin massless Lagrangians was recently proposed in~\cite{Dalmazi:2025fsl} where the gauge symmetry is also reduced with respect to the Fronsdal one. See also \cite{Campoleoni:2021blr}.} This property may just be a special feature of the free theory, or it may bear some additional significance.

 Via the Noether procedure, one can in principle reconstruct the full structure of a gauge theory, including both non-linear deformation of the symmetry and vertices  (up to curvature corrections) knowing its free sector only. Thus, the broader meaning of our observation should be sought in its possible relevance for higher-spin interactions (see e.g.~\cite{Bekaert:2022poo, Ponomarev:2022vjb} and references therein.). The issue at stake would be to see whether the consistent deformations of the reduced symmetry would amount only to those vertices that are known to be consistent for the full Fronsdal theory, or if new solutions would eventually emerge beyond the quadratic case. 

In the latter case, the putative new interactions derived from a reduced symmetry would break the full higher-spin symmetry of the free theory, which is expected to imply the appearance of unwanted degrees of freedom, or possibly a breakdown of perturbativity (see e.g.~\cite{Hell:2023mph,Karananas:2024hoh,Barker:2025gon}), thus putting into question the relevance of the mechanism we explored in this paper beyond the free theory. Differently, if only the consistent vertices emerge from this analysis, one would be led to speculate about the existence of some connection between non-linearly deformed higher-spin symmetries and diffeomorphisms, with the former emerging from the request of invariance under some high-derivative implementation of the latter.

\begin{acknowledgments}
This work used the DiRAC Data Intensive service~(CSD3 \href{www.csd3.cam.ac.uk}{www.csd3.cam.ac.uk}) at the University of Cambridge, managed by the University of Cambridge University Information Services on behalf of the STFC DiRAC HPC Facility~(\href{www.dirac.ac.uk}{www.dirac.ac.uk}). The DiRAC component of CSD3 at Cambridge was funded by BEIS, UKRI and STFC capital funding and STFC operations grants. DiRAC is part of the UKRI Digital Research Infrastructure.

This work also used the Newton compute server, access to which was provisioned by Will Handley using an ERC grant.

W.~B. is grateful for the support of Girton College, Cambridge, Marie Sklodowska–Curie Actions (MSCA) and the Institute of Physics of the Czech Academy of Sciences. The work of C.~M. was supported by the Estonian Research Council grant PRG1677. A.~S. acknowledges financial support from the ANID CONICYT-PFCHA/DoctoradoNacional/2020-21201387.

Co-funded by the European Union (Physics for Future – Grant Agreement No. 101081515). Views and opinions expressed are however those of the author(s) only and do not necessarily~reflect those of the European Union or European Research Executive Agency. Neither the European Union nor the granting authority can be held responsible for them.
\end{acknowledgments}

\appendix

%%%
\section{Notation and conventions} \label{notation}
%%%

When working with symmetric tensors of arbitrary rank it is useful to employ a notation where symmetrized indices are implicit, traces are denoted by `primes' or by a number in square brackets and other contractions of indices, whenever not ambiguous, by a `dot', e.g.
\begin{subequations}
\begin{align}
\vf &\MyEquiv \vf_{\m_1 \ldots \m_s}, \\
\vf^{\, \prime} &\MyEquiv \h^{\a \b} \vf_{\a \b \m_3 \ldots \m_s}, \\
\prd \vf &\MyEquiv \pr^{\a} \vf_{\a \m_2 \ldots \m_s}.
\end{align}
\end{subequations}
The resulting combinatorial rules are summarized in the following identities~\cite{Francia:2002aa, Francia:2007qt}
\begin{subequations}
\begin{align}
\left( \pr^{\, p} \, \vf \right)^{\, \pe} &= \Box \, \pr^{\, p-2} \, \vf + 2 \, \pr^{\, p-1} \, \prd \vf + \pr^{\, p} \, \vf^{\, \pe} , \\
\partial^{\, p} \, \partial^{\, q} &= \smallbinom{p+q}{p} \partial^{\, p+q} , \\
\partial \cdot \left( \partial^{\, p} \vf \right) &= \Box \, \partial^{\, p-1} \vf + \partial^{\, p} \partial \cdot \vf , \\
\partial \cdot \eta^{\, k} &= \partial \, \eta^{\, k-1} , \\
\left( \eta^k \, \vf \right)^{\, \prime} &= \Big[ D + 2 (s+k-1) \Big] \eta^{\, k-1} \, \vf + \eta^k \, \vf^{\, \prime} , \\
(\vf \, \psi)^{\, \pe} &= \vf^{\, \pe} \, \psi + \vf \, \psi^{\, \pe} + 2 \, \vf \cdot \psi , \\
\eta \, \eta^{\, n-1} &= n \, \eta^{\, n} ,
\end{align}
\end{subequations}
where symmetrizations involve the minimum number of terms and do not include normalization factors. 

%%%
\section{Failure of scalar reduction}\label{Failure}
%%%

The condition for gauge invariance given in~\cref{deltaEkorder0}, when written in full, entails separately vanishing expressions for each~$n$ corresponding to the~${\cal O} (\h^n)$ parts. Each of these expressions is linear in the~$\l_1$,~$\ldots$,~$\l_5$, with coefficients that depend on~$s$ and~$D$. To study the~$s$- and~$D$-dependence of the solutions systematically, it is prudent to make this linearity homogeneous, and so we introduce a parameter~$\l_0$ multiplying the~$\Box \vf$ term in~\cref{E_k}. Setting~$\l_0 \to 1$ in what follows is always safe, amounting to an innocuous field rescaling. There is then another subtlety in terms of structural~$s$-dependence. For~$s<3$ some of the operators in~\cref{E_k} vanish identically, so that the coefficients multiplying them do not enter the problem. We thus consider a variable-length column vector~$\mathsf{L}(s)$, whose elements contain~$\l_0$ for~$s\geq 0$, and~$\l_1$ for~$s\geq 1$, and~$\l_2$,~$\l_3$ and~$\l_5$ for~$s\geq 2$, and~$\l_4$ for~$s\geq 3$. Similarly, the integer~$n$ runs from zero to~$\lfloor s/2 \rfloor$, and we arrange the sorted~$n$ into the variable-length~$\mathsf{N}(s)$. With this notation, the scalar-reduced gauge invariance~$\d E_k = 0$ requires that the column vector obey~$\mathsf{V}(s,D)=0$, where~$\left[\mathsf{V}(s,D)\right]_i\MyEquiv V\big(\mathsf{L}(s),\left[\mathsf{N}(s)\right]_i|s,D\big)$ and we define
\begin{align}
	&
	V\big(\mathsf{L}(s\geq 3),n|s,D\big)\MyEquiv\frac{\a_{n+1}}{2}(s-2n)(s-2n-1)
	\nonumber\\
	&\hspace{20pt}
	\times(s-2n-2)\Big[
		(\l_2+2n\l_4)\big(D+2(s-n-2)\big)
	\nonumber\\
	&\hspace{10pt}
		+2(\l_1+n\l_5)
		\Big]
	+{\a_n}\Big[
		\l_0(s-2n)
		+\l_1(s-2n)^2
	\nonumber\\
	&\hspace{10pt}
	+\l_2(s-2n)\smallbinom{s-2n}{2}
		+\big(\l_3+(s-2n)\l_4\big)n(s-2n)
	\nonumber\\
	&\hspace{20pt}
		\times\big(D+2(s-n-1)\big)
		+\big(
			\l_4(s-2n)(s-2n-1)
	\nonumber\\
	&\hspace{10pt}
			+\l_5(s-2n+1+s-2n)
		\big)n(s-2n)
		\Big]
	\nonumber\\
	&
	+{\a_{n-1}}
		\Big[\l_3
		+(s-2n)\l_4
	+\l_5\Big]n(s-2n+2),\label{ScalarFunction}
\end{align}
with~$V\big(\mathsf{L}(s<3),n|s,D\big)$ defined by setting various parameters to zero in~\cref{ScalarFunction} according to the restrictions above. For all choices of~$s$ and~$D$, the conditions for gauge invariance can thus be expressed as the problem~$\mathsf{V}(s,D)=\mathsf{M}(s,D)\cdot\mathsf{L}(s)=0$ where~$\left[\mathsf{M}(s,D)\right]_{ij}\MyEquiv\partial V\big(\mathsf{L}(s),\left[\mathsf{N}(s)\right]_i|s,D\big)/\partial\left[\mathsf{L}(s)\right]_j$. This system will constrain the Lagrangian parameters according to its rank. The Fronsdal operator and its trace --- where defined --- account for~$\mathsf{f}(s)$ parameters, where~$\mathsf{f}(s\geq 2)\MyEquiv 2$ and~$\mathsf{f}(s<2)\MyEquiv 1$. Altogether, the scalar reduction yields~$N(s,D)$ free parameters beyond the Fronsdal solution, where
\begin{equation}\label{NCalc}
	N(s,D)\MyEquiv\operatorname{dim}\big(\mathsf{L}(s)\big)-\operatorname{rank}\big(\mathsf{M}(s,D)\big)-\mathsf{f}(s).
\end{equation}
The rank in~\cref{NCalc} corresponds to the number of non-zero singular values, i.e., the square roots of the eigenvalues of the positive semidefinite Gram matrix~$\mathsf{M}(s,D)^{\text{T}}\cdot\mathsf{M}(s,D)$. As shown in the supplement~\cite{Supplement}, the general result is that one singular value~$\sigma_1>0$ is present for~$s\geq 0$, with a further~$\sigma_2>0$ for~$s\geq 2$ and~$\sigma_3>0$ for~$s\geq 4$. No further non-zero singular values arise, and so for large~$s$ the system stabilizes with solutions
\begin{subequations}
\begin{align}
	\l_2 &= -\frac{\l_0(D+2s-6) + \l_1(s-2)(D+s-3)}{(s-1)(D+s-4)}, \\
	\l_4 &= -\l_5, \\
	\l_3 &= \frac{2\l_0 + 2\l_1 - \l_5(s-1)(D+s-4)}{2(s-1)(D+s-4)}.
\end{align}
\end{subequations}
Success of the scalar reduction in uniquely motivating the Fronsdal solution corresponds to~$N(s,D)=0$. Taking into account~\cref{NCalc} and the number of non-zero singular values, we thus find success of the scalar reduction for~$N(0,D)=N(1,D)=0$, and failure of the scalar reduction for~$s\geq 2$, specifically~$N(2,D)=1$,~$N(3,D)=2$, and~$N(s\geq 4,D)=1$. Even within this~$s$-dependent scheme, anomalies occur at~$D=1$ for~$s=3$,~$s=4$ and~$s=5$, coinciding with roots of the singular values. Anomalous system ranks are \emph{smaller} than expected, leading to a \emph{larger} loss of uniqueness for the Fronsdal solution. The root system of singular values is illustrated by taking the continuous analytic extension of~$s$ and~$D$ in~\cref{fig:singular_values_francia}.

\bibliography{Manuscript}

\end{document}